\documentclass{article}
\usepackage{arxiv}

\usepackage[utf8]{inputenc}
\usepackage[T1]{fontenc}
\usepackage{booktabs}
\usepackage{makecell}
\usepackage{amsmath}
\usepackage{amssymb}
\usepackage[exponent-product=\cdot]{siunitx}
\usepackage{subcaption}
\usepackage{graphicx}
\usepackage{physics}
\usepackage{derivative}
\usepackage{microtype}
\usepackage[numbers]{natbib}
\usepackage[hidelinks,hypertexnames=false]{hyperref}
\usepackage{url}
\usepackage{doi}
\usepackage{cleveref}

\title{A Topology-Based Approach to Representative Volume Element Assessment in Solid Oxide Cell Anodes Using Large Volume Tomography and Persistent Homology}

\author{%
  Aleksandra Kukiełka\textsuperscript{1},
  Maksym Szemer\textsuperscript{1,2},
  Szymon Buchaniec\textsuperscript{1,4}\\
  \bfseries Katarzyna Berent\textsuperscript{3},
  Grzegorz Brus\textsuperscript{1}\thanks{Corresponding author. \texttt{brus@agh.edu.pl}}\\[6pt]
  \begin{minipage}{0.96\textwidth}
    \centering\normalfont\footnotesize
    \textsuperscript{1}Department of Fundamental Research in Energy Engineering, Faculty of Energy and Fuels, AGH University of Krakow, 30 Mickiewicza Ave., Krakow, 30-059, Poland\\[3pt]
    \textsuperscript{2}Faculty of Electrical Engineering, Automatics, Computer Science and Biomedical Engineering, AGH University of Krakow, 30 Mickiewicza Ave., Krakow, 30-059, Poland\\[3pt]
    \textsuperscript{3}Academic Centre for Materials and Nanotechnology, AGH University of Krakow, 30 Mickiewicza Ave., Krakow, 30-059, Poland\\[3pt]
    \textsuperscript{4}Department of Mechanical Engineering and Science, Kyoto University, Nishikyo-ku, Kyoto, 615-8540, Japan
  \end{minipage}%
}
\date{}
\renewcommand{\shorttitle}{Representative Volume Element Assessment}
\hypersetup{
  pdftitle={A Topology-Based Approach to Representative Volume Element Assessment in Solid Oxide Cell Anodes Using Large Volume Tomography and Persistent Homology},
  pdfauthor={Aleksandra Kukiełka, Maksym Szemer, Szymon Buchaniec, Katarzyna Berent, Grzegorz Brus},
  pdfkeywords={SOFC, RVE, FIB-SEM, plasma-FIB, microstructure, computational materials}
}

\begin{document}
\maketitle

\begin{abstract}
Determining the representative volume element is essential for establishing statistically meaningful microstructure--property relationships in heterogeneous materials and porous media. Existing approaches rely on the convergence of geometric descriptors, such as porosity, tortuosity, or specific surface area, leading to descriptor-dependent estimates of the representative volume element. In this work, we present a topology-based framework for representative volume element assessment. The proposed methodology is demonstrated on nickel--yttria-stabilized zirconia solid oxide cell anode reconstructed using plasma focused ion beam--scanning electron microscopy, resulting in one of the largest reported three-dimensional datasets for this material (\SI{121.8}{\micro\metre} $\times$ \SI{83.97}{\micro\metre} $\times$ \SI{45.99}{\micro\metre}). Persistent homology was used to characterize the multiscale topology of the pore network, and persistence images were compared with those of the largest available reference volume by measuring the $L_1$ distance between persistence image vectors. The resulting topological distance decreases monotonically with increasing observation volume and is accurately described by exponential decay models over the investigated range ($R^2 \in [0.972,\,0.983]$), providing a continuous measure of representativeness. To complement the experimental data, the framework was additionally tested on three independent synthetic realizations of the same stochastic microstructure. The proposed approach provides a practical tool for quantifying the trade-off between observation volume and representativeness error and is readily applicable to other porous and multiphase microstructures exhibiting scale-dependent behavior.
\end{abstract}


\clearpage

\section{Introduction}
\label{sec:intro}
The representative volume element (RVE) is a key concept in the quantitative characterization of heterogeneous materials and the validation of numerical models. An RVE is commonly defined as the smallest volume that statistically represents the bulk material~\cite{Kanit.2003}. In porous and multiphase systems, such as solid oxide fuel cell (SOFC) electrodes, the determination of an appropriate RVE is essential to ensure that microstructural parameters derived from three-dimensional reconstructions reliably reflect the effective transport properties and electrochemical behavior of the electrode~\cite{Vafaeenezhad.2022}.

In most existing studies, representative volume element size is determined a posteriori by analyzing the evolution of selected microstructural descriptors as a function of the sampled volume. Typically, the analyzed volume is gradually increased, and the RVE is identified once a given property converges to a stable value or fluctuates within a predefined tolerance~\cite{Joos.2011, Laurencin.2012, Harris.2015}. Commonly used descriptors include phase volume fractions~\cite{Ruse.2024}, specific surface area, and the tortuosity factor~\cite{Harris.2015, Joos.2012}. Alternative approaches involve the comparison of averaged quantities extracted across independent regions of the microstructure~\cite{Ruse.2024, Epting.2017, Mahbub.2021}.

A key limitation of these approaches is that the estimated RVE size depends on the specific descriptor selected for the analysis. Different geometric or transport-related quantities may converge at different length scales, meaning that the RVE obtained for one descriptor can differ from that obtained for another, even for the same microstructure. Moreover, conventional descriptors typically characterize individual structural properties, such as phase fraction or interfacial area, and may not fully capture the organization of complex porous networks, including their connectivity and topology.

The reliability of representative volume element determination also strongly depends on the available imaging data. Conventional focused ion beam-scanning electron microscopy (FIB-SEM) has been widely used for three-dimensional reconstruction of SOFC electrodes~\cite{ Sciazko.2023, Zhang.2023, Hoffrogge.2023}. Due to its high spatial resolution, with voxel sizes typically ranging from \SIrange{10}{100}{\nm}~\cite{Wilson.2011}, FIB-SEM provides detailed microstructural information. However, reconstructed volumes are often limited in size (typically \qtyproduct[product-units = single]{10 x 10 x 10}{\micro\meter\cubed}, with one of the largest reported volumes being \qtyproduct[product-units = single]{45.8 x 14.5 x 26.2}{\micro\meter\cubed}~\cite{Kanno.2011}), which can restrict the accessible observation scale and prevent reliable identification of the representative regime. Synchrotron X-ray nanotomography enables reconstruction of substantially larger volumes (tens to hundreds of micrometers per side), but generally involves a trade-off between accessible volume and spatial resolution, with voxel sizes commonly ranging from \SI{35}{\nm} to \SI{150}{\nm}~\cite{Zekri.2017}. This limitation is particularly important for materials with complex and heterogeneous pore networks, where large observation volumes may be required to capture representative structural organization.

Recent advances in plasma FIB-SEM (PFIB-SEM) tomography enable reconstruction of substantially larger volumes while maintaining nanoscale spatial resolution, with reported datasets reaching \qtyproduct[product-units = single]{750 x 143 x 310}{\micro\meter\cubed} for Li-ion battery cathodes~\cite{Yoshida.2024} and \qtyproduct[product-units = single]{126 x 73 x 12.5}{\micro\meter\cubed} and \qtyproduct[product-units = single]{124 x 110 x 8}{\micro\meter\cubed} for commercial SOFC cathodes and anodes~\cite{Hsu.2018}. These larger datasets enable investigation of microstructural convergence over a broader range of observation scales and substantially increase the likelihood of reaching the representative scale.

However, increasing the reconstructed volume alone does not resolve the fundamental limitation associated with descriptor-dependent RVE estimation. Even selecting the largest RVE obtained from several conventional descriptors cannot guarantee overall representativeness, since additional microstructural descriptors may converge only at larger observation scales. From a practical perspective, the observation volume should be large enough to capture the characteristic microstructural organization, while remaining as small as possible to limit the computational cost of image segmentation, three-dimensional reconstruction, and subsequent microstructural analysis or numerical modeling. A criterion that captures complementary aspects of microstructural organization, rather than relying on a single selected geometric quantity, is therefore required.

To address this challenge, we employ topological data analysis (TDA), in particular, persistent homology (PH), for RVE determination of SOFC anodes. Persistent homology quantifies the evolution of connected components, tunnels, and cavities across a range of filtration scales, providing a multiscale representation of microstructural organization. Unlike conventional geometric descriptors, persistent homology characterizes the structural arrangement of the pore network, including complementary information related to connectivity and morphology.

Persistent homology has recently emerged as a versatile tool for microstructure characterization in materials science, with applications ranging from metallic alloys~\cite{Wang.2020} and porous rocks~\cite{Herring.2018} to pathology detection in porous bone tissue~\cite{Pritchard.2023} and heterogeneous media topology-aware design~\cite{Gao.2024}. PH has also been applied to the analysis of solid oxide fuel cell electrode microstructures~\cite{Pawowski2023}. Its potential for detailed microstructure characterization and for enhancing data-driven SOFC modeling has also been recognized~\cite{Brus.2023}, leading to applications including the validation of numerical models of SOFC anodes~\cite{Yamatoko.2025}, prediction of current-voltage performance using artificial neural networks~\cite{Szemer.2025}, and prediction of microstructural parameters derived from topological descriptors~\cite{Szemer.2026}. These studies demonstrate the potential of persistent homology for characterizing complex electrochemical microstructures. However, its application for determining the representative volume element has not yet been established.

In this work, we propose a topology-based framework for the representative volume element assessment of a Ni-YSZ SOFC anode. Persistence images are used to encode the topology of pore networks across increasing observation volumes, and their similarity is quantified by the distance between each subvolume and the largest available volume used as a reference. This enables RVE assessment to be formulated as a quantitative convergence problem in a topological feature space, with the resulting distance serving as a continuous measure of representativeness error. The framework is demonstrated for the three homology dimensions of the pore network and provides a systematic basis for assessing the observation scale required for microstructural representativeness.
\section{Research Methodology}
\label{sec:method}
\subsection{Experimental method}
\subsubsection{Plasma FIB-SEM} \label{subsec:FIB-SEM}
Large-volume three-dimensional reconstruction was performed using a Thermo Fisher Scientific Helios 5 UXe plasma-focused ion beam scanning electron microscope (PFIB-SEM). The SOFC sample was embedded in Struers EpoFix epoxy resin and left to cure overnight at room temperature. To remove trapped air bubbles and ensure complete infiltration of the porous SOFC ceramic structure, vacuum impregnation was performed using a Struers CitoVac system. After curing, a fragment of the SOFC cross-section was cut out and mechanically polished using 4000-grit SiC abrasive paper. The samples were cleaned with isopropanol. The specimen was mounted on an aluminum SEM stub using silver paint and coated with a \SI{60}{\nano\metre} thick carbon layer using a Q150T ES Quorum Technologies sputter coater. For PFIB-SEM measurements, the sample was tilted to \SI{52}{\degree}, corresponding to a position perpendicular to the ion beam. A region of interest (ROI) with dimensions of \qtyproduct{210x120}{\micro\m} was selected. Prior to serial sectioning, the ROI surface was prepared using the cleaning cross-section procedure to minimize curtaining effects and improve surface quality.
The ROI was first protected with a thin tungsten (W) layer deposited by electron-beam-induced deposition (EBID), with an estimated thickness of approximately \SI{200}{\nano\metre}. Subsequently, a \SI{5}{\micro\metre} thick protective W layer was deposited by ion-beam-induced deposition (IBID) using a \SI{30}{\kilo\volt} accelerating voltage and a beam current of \SI{7.0}{\nano\ampere}.
Serial sectioning and image acquisition were performed automatically using AutoSlice and View software (version 5.10). A slice thickness of \SI{70}{\nano\metre} was used throughout the experiment, corresponding to the voxel dimension in the $z$-direction. A total of 675 slices were acquired using Xe$^{-}$ ions at \SI{30}{\kilo\volt} and \SI{60}{\nano\ampere}.
SEM images were collected using a Through-the-Lens Detector for Secondary Electrons (TLD-SE) at an accelerating voltage of \SI{2}{\kilo\volt}, a beam current of \SI{400}{\pico\ampere}, and a working distance of \SI{4}{\milli\metre}. Images were acquired at a resolution of \qtyproduct{6417x5740}{pixels} with a dwell time of \SI{5}{\micro\second} per pixel, resulting in a total imaging time of 3 min 4 s per slice. Image contrast and brightness were adjusted prior to acquisition and kept constant throughout the automated run without software modification. Images were stored in 8-bit format. The voxel size of the reconstructed PFIB-SEM dataset was \qtyproduct{30x30x70}{\nano\metre} ($x-y-z$).

\subsubsection{Segmentation and 3D reconstruction} \label{subsec:Segmentation}

The raw tomographic volume acquired by PFIB-SEM measured \qtyproduct{192.48 x 172.17 x 47.04}{\micro\metre} at a native voxel size of \qtyproduct{30x30x70}{\nano\metre}. The dataset was imported into Avizo 3D software (Thermo Fisher Scientific, version 2024.2) and subjected to slice-by-slice alignment, after which the volume was cropped to \qtyproduct{121.8 x 147.9 x 45.99}{\micro\metre}. The cropped volume was then visually inspected to identify regions compromised by PFIB-SEM acquisition artifacts. Slices exhibiting the curtain effect, caused by non-uniform ion beam milling, and regions containing incompletely epoxy-filled pores introduced during sample preparation, were excluded. The resulting subvolume of \qtyproduct{121.8 x 83.97 x 45.99}{\micro\metre} was retained for segmentation.
Segmentation of the pore phase was performed using the AI Assisted Segmentation tool within the Avizo Segmentation+ workroom \cite{chen.2017}. A sparse manual annotation was generated with the brush tool on selected cross-sections. To ensure that the training data captured the morphological variability of the pore network across the full spatial extent of the volume, 9 regions of interest (ROIs) were defined at systematically varied locations, including front, back, middle, top, and bottom positions within the dataset. A VGG16 convolutional neural network \cite{Simonyan.2014} was selected as the segmentation model and trained on the annotated ROIs. The trained model was subsequently applied to the entire volume using the Compute Label function. No post-processing filters were applied to the resulting label image. To correct for the anisotropic native voxel geometry, the segmented volume was resampled to an isotropic voxel size of \qtyproduct{30 x 30 x 30}{\nano\metre} prior to microstructural analysis.

\begin{figure}[htbp]
    \centering
    \begin{subfigure}[b]{0.49\textwidth}
        \centering
        \includegraphics[width=\textwidth]{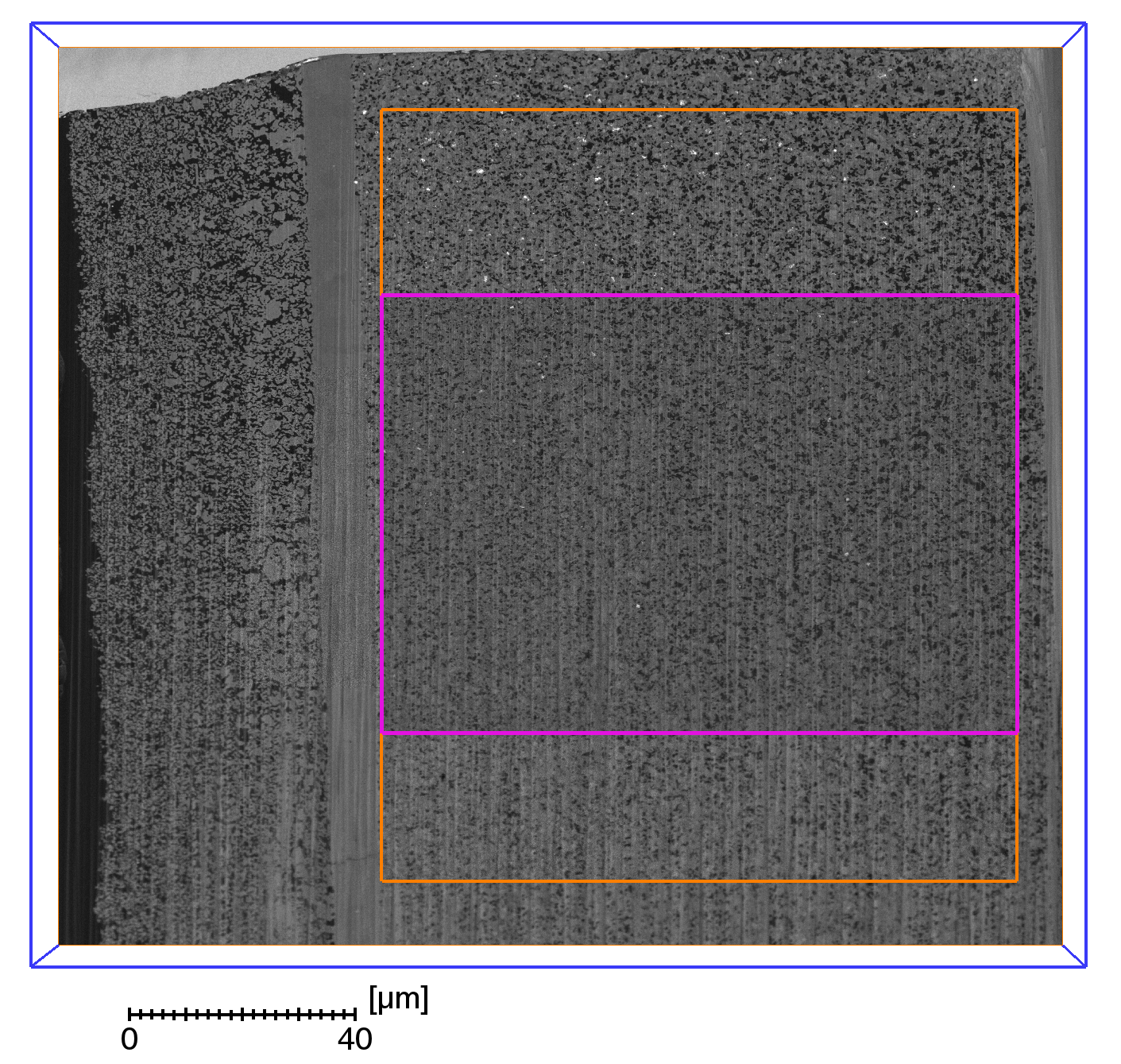}
        \caption{Region of interest and its location in anode-electrolyte-cathode assembly }
        \label{fig:roi}
    \end{subfigure}
    \hfill 
    \begin{subfigure}[b]{0.49\textwidth}
        \centering
        \includegraphics[width=\textwidth]{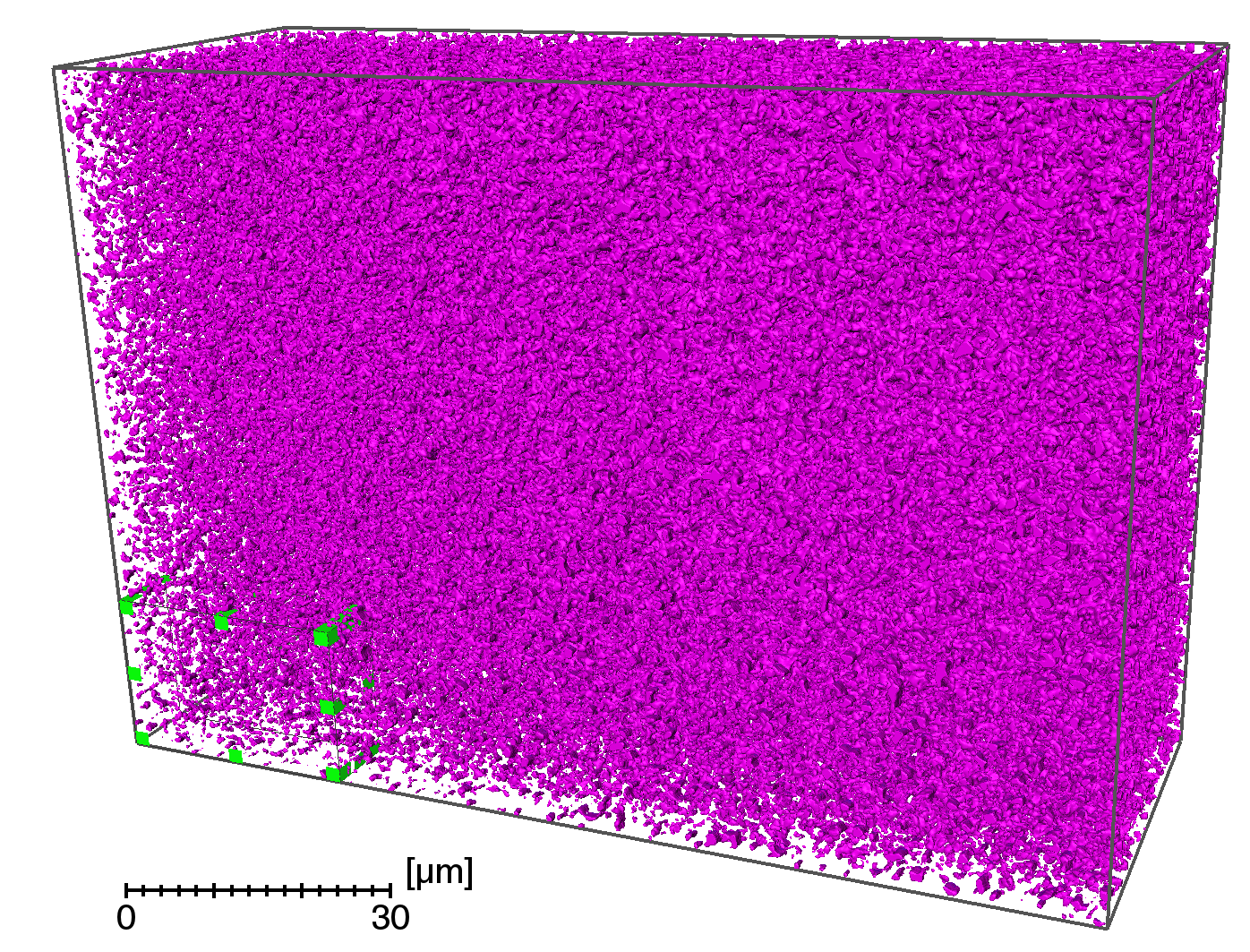}
        \caption{Digital material representation of porous structure in the investigated anode material}
        \label{fig:dmr}
    \end{subfigure}
    
    \caption{Selected region of interest and three-dimensional digital material representation of investigated volume element a) a cross-section of the  investigated volume, frames indicate measured cross-section (blue), after alignment and cropping (orange), and region selected for segmentation (magenta) b) Digital material representation of porous structure in the investigated anode material, a green frame represents a typical FIB-SEM  volume size}
    \label{fig:combined_figures}
\end{figure}


\subsection{Topology-based representation and computational framework}
\label{subsec:Computational method}

Prior to the topological analysis, the homogeneity of the reconstructed microstructure was evaluated using slice-by-slice porosity profiles calculated along the three principal axes of the full reconstructed volume (\qtyproduct[product-units = single]{4061 x 2800 x 1536}{voxels}) -- see~\cref{subsec:homogeneity}. Based on these distributions, a region exhibiting approximately stable porosity was selected for subsequent RVE analysis. All classical and topological descriptors described below were evaluated within this region.

Classical RVE estimation was first performed using porosity and specific surface area as reference geometric descriptors. Porosity was calculated as the fraction of voxels classified as pore phase within each subvolume. Specific surface area $S_v$ was determined using the marching cubes algorithm implemented in the \texttt{scikit-image} library \cite{skimage.2014} (version 0.26.0), which generates a triangulated representation of the pore-solid interface. These descriptors provide a conventional baseline for comparison with the proposed topology-based approach.

To investigate scale-dependent topological convergence, a set of 100 nested cuboidal subvolumes was extracted from the selected region. All subvolumes shared the same origin point, located adjacent to the electrolyte-anode interface, and were generated by uniformly scaling the dimensions while preserving the aspect ratio corresponding to the investigated region. This approach ensured that the spatial character of the microstructure was sampled consistently across all observation scales. The subvolume dimensions were gradually increased up to the full selected region, which served as the reference structure corresponding to the maximum available observation scale.

\subsubsection{Persistent homology-based representation}

Topology provides a description of microstructure organization that is complementary to classical geometric descriptors. While quantities such as porosity or specific surface area characterize individual geometric properties, topology describes how structural elements are connected and arranged. For three-dimensional porous materials, the relevant topological features correspond to connected components ($H_0$), loops or tunnels ($H_1$), and enclosed cavities or voids ($H_2$). Persistent homology extends this concept by tracking the appearance and disappearance of such features across a range of filtration scales, providing a multiscale representation of microstructural organization~\cite{Edelsbrunner.2008}.

For voxelized tomographic data, persistent homology is naturally represented using cubical complexes, where voxels and their lower-dimensional faces form the elementary cells of the structure. Instead of analysing a single static representation, a filtration is generated by progressively including regions of the microstructure according to a scalar function. In this work, the filtration function was defined using the signed Manhattan distance to the phase boundary.

The reconstructed volume was segmented into two phases: pore and solid (Ni-YSZ combined). Since the present study focuses on pore-network topology, persistent homology was evaluated for the pore phase. For each subvolume, the signed Manhattan distance function was calculated as:
\begin{equation}
f_P(x)=
\begin{cases}
-\operatorname{dist}_1(x, \varGamma_\varOmega), & x\in \varOmega,\\
\operatorname{dist}_1(x,\varGamma_\varOmega), & x\notin \varOmega ,
\end{cases}
\label{eq:Manhattan_dist}
\end{equation}
where $\varOmega$ denotes the set of voxels assigned to the analyzed phase, $\varGamma_\varOmega$ represents the interface between the pore and solid phases, and $\mathrm{dist}_1(x,\varGamma_\varOmega)$ is the Manhattan distance between the voxel $x$ and boundary $\varGamma_\varOmega$. The signed Manhattan distance field was used as the filtration function. By assigning scalar values to the vertices of the cubical complex and extending them to higher-dimensional cells using the upper--star filtration rule~\cite{DeyWang.2022}, a nested sequence of complexes was generated in which topological features appear and disappear as the filtration parameter changes, enabling their tracking through persistent homology.

Persistent homology in dimensions $k=0,1,2$ was computed using the \texttt{cubicle} library \cite{Wagner.2023.cubicle} (version v0.8beta), producing sets of birth-death pairs:
\begin{equation}
D_k=\{(b_i,d_i)\in\mathbb{R}^2:b_i<d_i\}.
\label{eq:PD}
\end{equation}
Each point represents a topological feature ( $H_0$ -- isolated components, $H_1$ -- closed loops or $H_2$ -- enclosed voids), where $b_i$ and $d_i$ correspond to the filtration values at which the feature appears (birth) and disappears (death), respectively. The plot of this data in birth-death space is referred to as a persistence diagram (PD). Points close to the diagonal in PD correspond to features with short persistence, which may represent noise or fine-scale structural variations, whereas points farther from the diagonal represent more persistent topological characteristics.

Persistence diagrams were converted into persistence images (PIs) to obtain fixed-size numerical representations suitable for quantitative comparison. A common axis range was imposed across all persistence diagrams. Global extrema $(b_{\min}, b_{\max})$ and $(d_{\min}, d_{\max})$ were determined over the entire dataset and used to define a shared birth-death domain for all PI diagrams. The implementation used in this work was performed directly in birth-death coordinates using the \texttt{HomCloud} library \cite{Obayashi.2022.HomCloud} (version 5.1.0). Persistence surfaces $\rho_k$ were obtained by representing each persistence point from PDs with a weighted Gaussian distribution~\cite{Adams.2017}:
\begin{equation}
\rho_k(\mathbf{x})=
\sum_{\mathbf{u_i}}
w(\mathbf{u_i})
\frac{1}{2\pi\sigma^2}
\exp
\left(
-\frac{\|\mathbf{x}-\mathbf{u_i}\|^2}{2\sigma^2}
\right),
\label{eq:PI_sum}
\end{equation}
where $k=0, 1, 2$ is homology dimension, $\mathbf{x}=(b,d)$ denotes a point in the birth--death plane, $\mathbf{u_i}=(b_i,d_i)$ is the $i$-th persistence point, $w(\mathbf{u_i})$ is a weighting function, and $\sigma$ is the standard deviation of the Gaussian kernel controlling the degree of smoothing. The persistence surface is discretized on a regular grid to obtain the persistence image:
\begin{equation}
PI_k(i,j)=
\int_{P_{i,j}}\rho_k(x,y)\,\odif{x}\odif{y} .
\label{eq:PI}
\end{equation}
where $P_{i,j} \in [b_{\min}, b_{\max}] \times [d_{\min}, d_{\max}]$ is the grid region. The arctangent weighting function proposed by Kusano \textit{et al.} \cite{Kusano.2017} was applied:
\begin{equation}
w(b,d)=\arctan(c(d-b)^p),
\end{equation}
with $c=10^{-2}$ and $p=2$. This weighting increases the contribution of long-lived features while reducing the influence of short-lived features near the diagonal. A Gaussian kernel with $\sigma=1.0$ and a grid resolution of $n_{\mathrm{bins}}=30$ were used. PI hyperparameters were chosen following~\cite{Kusano.2017, Pawowski2023} and initial sensitivity analysis by Szemer \textit{et al.} ~\cite{Szemer.2025}. The resulting PI vectors were normalized by dividing by their $\ell^1$ norm (integral of $\rho_k$ over the persistence domain) before further analysis. The resulting persistence image provides a normalized representation of the topological feature distribution in the persistence domain.

\subsubsection{Topological convergence criterion}

The convergence of topological descriptors was quantified by calculating distance between the persistence image vector of each subvolume and that of the largest available reference volume:
\begin{equation}
D_i =
\norm{\vb{PI}_i-\vb{PI}_{\mathrm{ref}}}_1 ,
\label{eq:pi_dist}
\end{equation}
where $\vb{PI}_i$ denotes the persistence image vector of the $i$-th subvolume and $\vb{PI}_{\mathrm{ref}}$ corresponds to persistence image of the reference volume. The largest subvolume was selected as the reference because it represents the maximum available sampling volume and therefore provides the closest approximation of the bulk microstructural topology within the reconstructed dataset.

The resulting distance-volume curves provide a continuous measure of structural convergence, where decreasing distance indicates increasing similarity to the reference microstructure. For this reason, we interpret the topological distance as a measure of representativeness error. Since all subvolumes preserve a fixed aspect ratio, we use the length of the subvolume's shortest edge, $L$, as the characteristic length. To describe the decay of the PI distance with increasing subvolume size $L$, two candidate empirical functions were fitted to the distance-volume curves -- exponential:
\begin{equation}
    \hat{D}_{\exp}(L) = a \cdot \exp(-b L),
    \label{eq:exp_fit}
\end{equation}
and power-law:
\begin{equation}
    \hat{D}_{\mathrm{pow}}(L) = a \cdot L^{-n},
    \label{eq:pow_fit}
\end{equation}
where $a$, $b$ and $n$ are fit parameters. Fits were restricted to the range of \SIrange{10}{90}{\percent} of the investigated edge lengths ($L \approx \SIrange{3}{27}{\micro\metre}$): the lower bound excludes subvolumes comparable to the average particle size, and the upper bound excludes subvolumes that nearly coincide with the reference volume, for which $D \to 0$ by construction. Both fits and goodness-of-fit metrics were computed in log-space (i.e., on $\ln (D)$ as a function of $L$ for the exponential model, and on $\ln (D)$ as a function of $\ln(L)$ for the power-law model), to ensure appropriate weighting across the several orders of magnitude spanned by the PI distance. The fits were compared using the coefficient of determination ($R^2$) and the Bayesian information criterion (BIC). Based on this comparison, the exponential decay function $\hat{D}_{\exp}$ was selected for further analysis. The fitted model establishes a direct relationship between observation volume and topological representativeness error, allowing the minimum observation scale required to achieve a prescribed level of topological representativeness to be determined analytically. This formulation replaces the binary representative/non-representative classification with a continuous, quantitative measure of representativeness error.

\subsubsection{Validation on synthetic microstructures}
For additional validation of the proposed approach, three synthetic porous microstructures were generated using the random Gaussian fields method, following the approach established by Moussaoui et al.~\cite{Moussaoui.2018}. The samples were generated as three realizations of the same stochastic two-phase microstructure, with a volume fraction of \SI{30}{\percent} and correlation length $\ell = \SI{0.4}{\micro\metre}$. A grid spacing of $dx=\SI{0.030}{\micro\metre}$ was used, with different random seed values ($seed = 0, 1, 2$). Each synthetic sample consisted of \qtyproduct[product-units = single]{1000 x 1000 x 1500}{voxels}, corresponding to the dimensions of the quasi-homogeneous region selected from the experimental sample based on the porosity profiles (see \cref{subsec:homogeneity}).

Following the procedure applied to the experimental microstructure, each synthetic sample was divided into 100 nested subvolumes with an aspect ratio of 2:2:3. Persistence images were calculated for the resulting subvolumes using the same parameters as for the experimental data. The distances between the persistence images were evaluated using two reference schemes. First, for each synthetic realization, the largest subvolume was used as the self-reference, and the distances of the remaining subvolumes were calculated with respect to this reference. Second, the largest subvolume of one synthetic sample was used as a reference for the two other synthetic samples. This allowed the distances between subvolumes from different realizations of the same stochastic microstructure to be compared.
\section{Results and Discussion}
\label{sec:results}

\subsection{Classical RVE estimation}
\label{subsec:Classical_RVE}
\begin{figure}[htbp]
    \centering

    \begin{subfigure}[b]{0.48\textwidth}
        \centering
        \includegraphics[width=\textwidth]{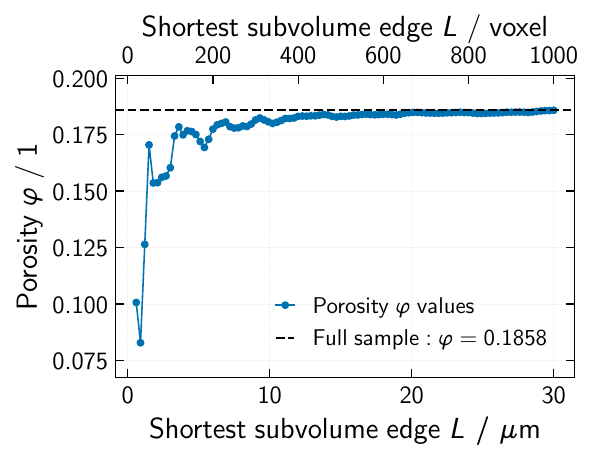}
        \caption{Porosity $\varphi$}
        \label{fig:rve_porosity}
    \end{subfigure}
    \hfill
    \begin{subfigure}[b]{0.48\textwidth}
        \centering
        \includegraphics[width=\textwidth]{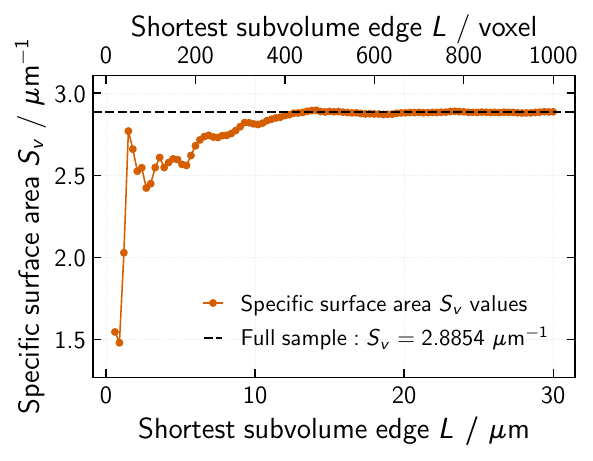}
        \caption{Specific surface area $S_v$}
        \label{fig:rve_sv}
    \end{subfigure}

    \caption{Classical RVE estimation based on (a) porosity $\varphi$ and (b) specific surface area $S_v$ as a function of subvolume size, expressed as the shortest edge $L$ of each nested cuboid subvolume. The dashed line indicates the reference value computed for the full homogeneous region (\qtyproduct[product-units = single]{1000 x 1000 x 1500}{voxels})}
    \label{fig:classical_RVE}
\end{figure}

Classical representative volume element analysis was first performed for the homogeneous region identified in \cref{subsec:homogeneity} using two commonly employed geometric descriptors: porosity ($\varphi$) and specific surface area ($S_v$). Both quantities were evaluated for 100 nested cuboidal subvolumes (see \cref{fig:classical_RVE}) generated with a fixed aspect ratio of 2:2:3, matching that of the reference volume. The subvolumes ranged from \qtyproduct[product-units = single]{10 x 10 x 15}{voxels} to the region of \qtyproduct[product-units = single]{1000 x 1000 x 1500}{voxels}, which served as the reference volume. Maintaining a constant aspect ratio preserves geometric self-similarity across observation scales and eliminates shape-dependent effects that could otherwise influence the calculated descriptors.

The two descriptors exhibit distinctly different convergence behavior. Specific surface area $S_v$ (\cref{fig:rve_sv}) exhibits fluctuations for small subvolumes, including a pronounced local minimum below \SI{2}{\micro\metre}, where the sampled volume is too small to be representative and boundary effects dominate. Beyond approximately \SI{15}{\micro\metre}, $S_v$ oscillates around the full-sample reference of \SI{2.885}{\micro\metre^{-1}} without showing a clear monotonic approach or a well-defined plateau. While the values remain close to the reference, the persistent fluctuations make it difficult to identify a specific subvolume size at which convergence can be confidently declared - one could argue for convergence anywhere beyond \SI{15}{\micro\metre}, depending on the accepted oscillation margin. Porosity (\cref{fig:rve_porosity}) follows a different trend. After the boundary-dominated regime below approximately \SI{2}{\micro\metre}, the porosity increases gradually throughout the investigated range without reaching a stationary value. This behavior is consistent with the residual porosity gradient along the electrode thickness identified in \cref{subsec:homogeneity}, indicating that even within the selected reference volume the pore volume fraction remains weakly location-dependent.

These results reveal two practical limitations of conventional representative volume element estimation. First, different descriptors exhibit fundamentally different convergence behavior. While $S_v$ approaches its reference value through oscillations, $\varphi$ continues to increase monotonically over the entire investigated range. Secondly, the classical descriptors indicate that representative behavior is fully achieved only for observation volumes approaching the reference volume size. Although the oscillations of $S_v$ become relatively small beyond approximately \SI{15}{\micro\metre}, porosity does not exhibit an unambiguous plateau even for the largest analyzed subvolume (\qtyproduct{30 x 30 x 45}{\micro\metre}). Notably, this reference volume already exceeds the dimensions of the largest conventional FIB-SEM reconstructions reported for SOFC anodes. This observation highlights the importance of sufficient observation volume when assessing representativeness from tomographic reconstructions.

These limitations motivate the topology-based framework presented in the following section. Rather than defining representativeness through descriptor-specific thresholds, the proposed approach quantifies structural convergence as a continuous measure of topological similarity to the largest available reference volume.

\subsection{Topology-based RVE estimation}

\begin{figure}[htbp]
    \centering
    \includegraphics[
    width=0.9\linewidth,
    height=0.70\textheight,
    keepaspectratio
]{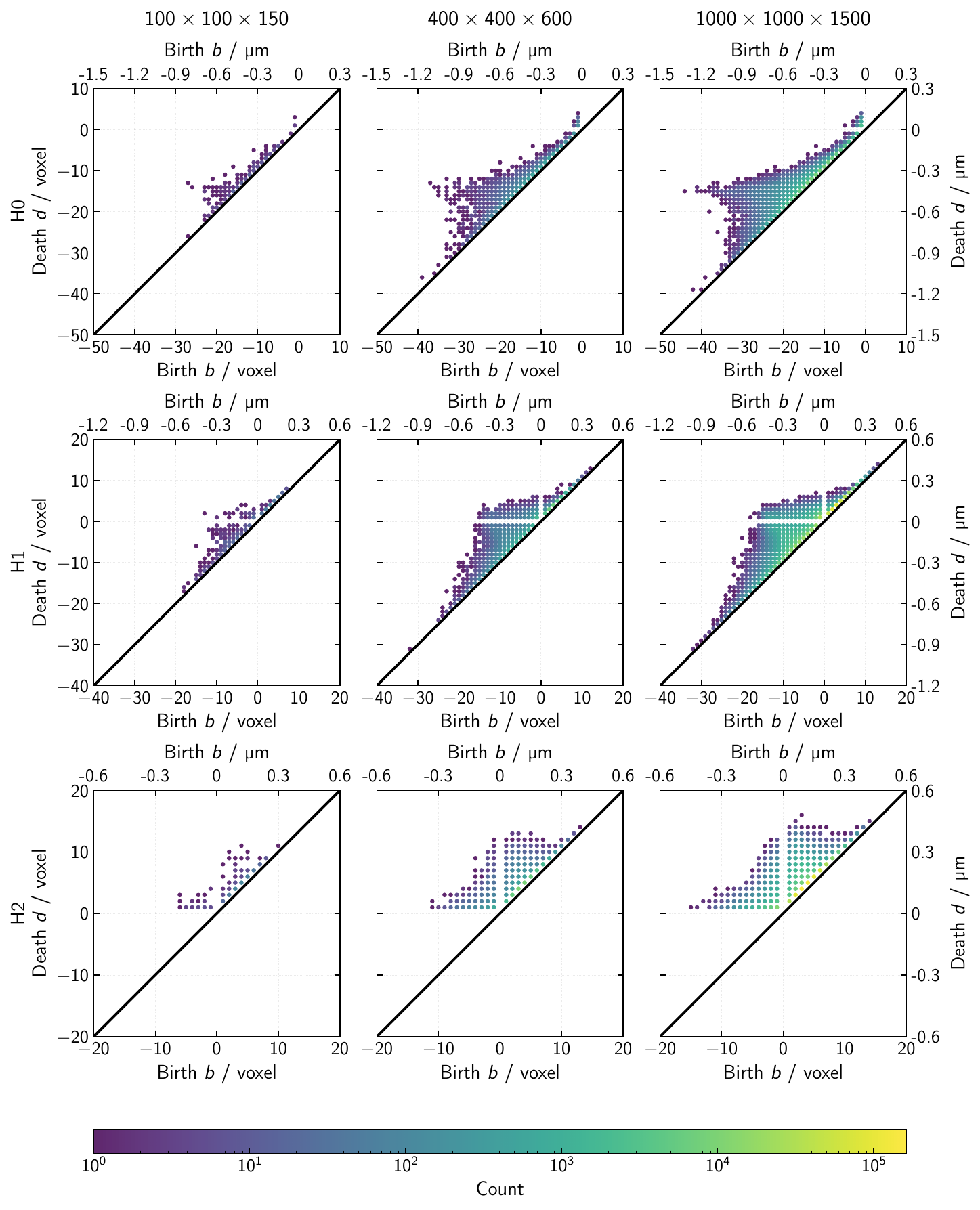}
    \caption{Persistence diagrams (PDs) for homology dimensions $H_0$, $H_1$, and $H_2$ computed for three subvolume sizes:\qtyproduct[product-units = single]{100 x 100 x 150}{voxels}, \qtyproduct[product-units = single]{400 x 400 x 600}{voxels}, and \qtyproduct[product-units = single]{1000 x 1000 x 1500}{voxels}. Rows correspond to homology dimensions and columns to increasing subvolume size. All diagrams within each homology dimension share identical axis ranges}
    \label{fig:PDs}
\end{figure}

\begin{figure}[htbp]
    \centering
      \includegraphics[
       width=0.8\linewidth,
       height=0.70\textheight,
       keepaspectratio
   ]{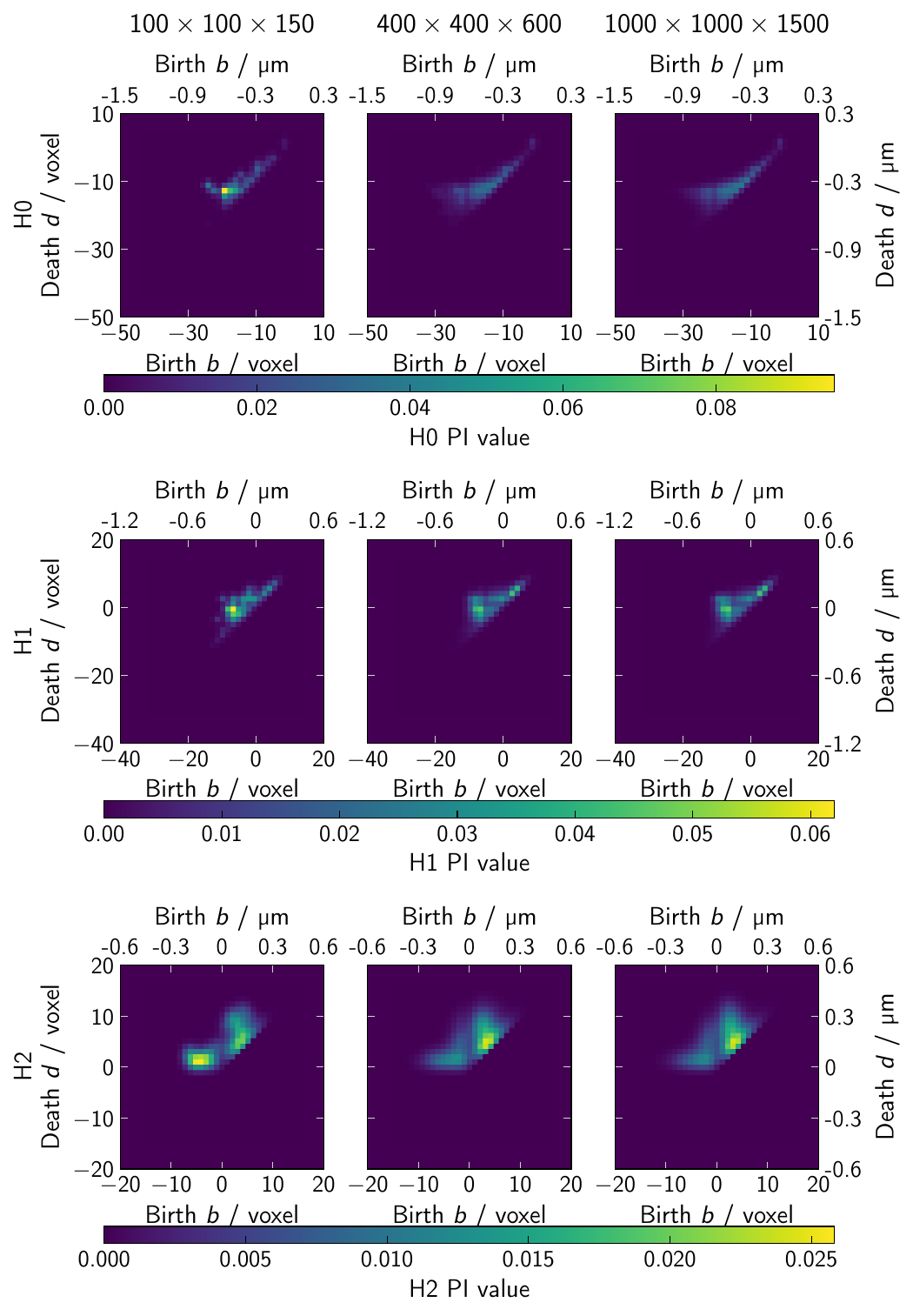}
    \caption{Persistence image (PI) representations in the $(b,d)$ plane corresponding to the persistence diagrams shown in \cref{fig:PDs} for homology dimensions $H_0$, $H_1$, and $H_2$. Rows correspond to the homology dimension, and columns correspond to increasing subvolume size }
    \label{fig:PIs}
\end{figure}

To address the ambiguity of classical representative volume element estimation, the same set of 100 nested subvolumes as in \cref{subsec:Classical_RVE} was analyzed using persistent homology. Three example subvolumes were selected to illustrate the evolution of the topological descriptors with increasing observation volume: small (\qtyproduct[product-units = single]{100 x 100 x 150}{voxels}), medium (\qtyproduct[product-units = single]{400 x 400 x 600}{voxels}), and large (\qtyproduct[product-units = single]{1000 x 1000 x 1500}{voxels}). Their persistence diagrams for homology dimensions $H_0$, $H_1$, and $H_2$ are presented in \cref{fig:PDs}. For the smallest subvolume, relatively few topological features are detected, with most points located close to the diagonal, indicating low-persistence structures. Increasing the observation volume produces more persistent features located farther from the diagonal, reflecting larger-scale structural characteristics of the pore network. Notably, the persistence diagrams of the medium and large subvolumes exhibit very similar point distributions, indicating that the overall topological structure has largely stabilized. The same trend is observed consistently for $H_0$, $H_1$, and $H_2$. To facilitate quantitative comparison, persistence diagrams were transformed into persistence images (PIs). Examples of PIs are shown in \cref{fig:PIs}. Consistent with the persistence diagrams, the smallest subvolume differs noticeably from the larger ones, whereas the medium and large subvolumes show a high degree of visual similarity, indicating stabilization of the distribution of persistent topological features with increasing observation volume.

\begin{figure}[htbp]
    \centering
    \begin{subfigure}[b]{0.49\textwidth}
        \centering
        \includegraphics[width=\linewidth]{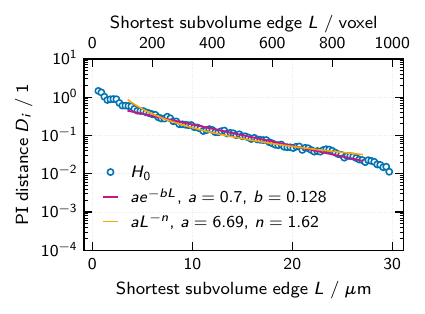}
        \caption{$H_0$}
        \label{fig:H0_fit}
    \end{subfigure}
    \hfill
    \begin{subfigure}[b]{0.49\textwidth}
        \centering
        \includegraphics[width=\linewidth]{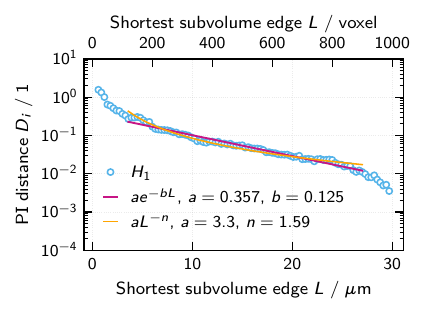}
        \caption{$H_1$}
        \label{fig:H1_fit}
    \end{subfigure}
    \par\medskip
    \begin{subfigure}[b]{0.49\textwidth}
        \centering
        \includegraphics[width=\linewidth]{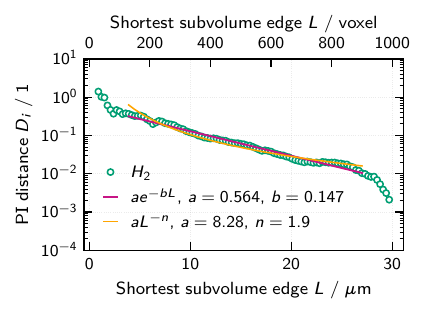}
        \caption{$H_2$}
        \label{fig:H2_fit}
    \end{subfigure}
    \caption{Distance between the persistence image (PI) of each subvolume and that of the reference volume as a function of subvolume size for homology dimensions $H_0$, $H_1$, and $H_2$. Exponential and power-law fits are shown together with the calculated distances. The vertical axis is plotted on a logarithmic scale}
    \label{fig:PI_distance_fit}
\end{figure}

The distance between the persistence image vector of each subvolume and that of the reference volume was calculated for each homology dimension (\cref{fig:PI_distance_fit}). In all cases, the distance decreases approximately monotonically with increasing observation volume, demonstrating progressive convergence toward the topology of the reference volume.

To quantify the observed decrease in topological distance, the distance-size curves were approximated using an exponential decay model: $\hat{D}_{\exp}(L) = a \cdot \exp(-bL)$. During fitting, subvolumes with an edge length below \SI{3}{\micro\metre} were excluded, since this scale is comparable to the average particle size of \SI{1}{\micro\metre}; at such small volumes, strong size effects emerge, and the extracted structure reflects the properties of individual grains rather than the overall microstructure. Subvolumes above \SI{27}{\micro\metre} (upper \SI{10}{\percent}) were also excluded, as their large overlap with the reference volume makes the distance vanish by construction (see \cref{fig:PI_distance_fit}, where the $H_1$ and $H_2$ curves deviate from the trend in this range).The fitting results are summarized in \cref{tab:fit_comparison}. The  exponential model provided high goodness of fit, with $R^2 \geq 0.972$ for all homology dimensions. The fitted decay rates were of similar magnitude across the three homology dimensions, with $b$ from \numrange{0.125}{0.147} (see \cref{tab:fit_comparison}).

The exponential form was compared with an alternative power-law model using both $R^2$ and the Bayesian information criterion (BIC). The exponential model yielded lower BIC values for all three homology dimensions, with substantially lower values than those obtained for the power-law model (\cref{tab:fit_comparison}). The same preference is observed in the corresponding $R^2$ values. These results indicate that, within the investigated observation-size range and for the experimental microstructure, exponential decay provides a better empirical description of the observed PI-distance convergence than the power-law model.

 \begin{table}[htbp]
   \centering
   \footnotesize
   \setlength{\tabcolsep}{4pt}
\caption{Parameters, $R^2$ and BIC values of the exponential ($\hat{D}_{\mathrm{exp}} = a\,\exp(-bL)$) and power-law ($\hat{D}_{\mathrm{pow}} = a\,L^{-n}$) fits applied to the PI distance curves for individual homology dimensions $H_{\mathrm{dim}}$. Fits, $R^2$ and BIC computed in log-space. Values in parentheses are standard errors of the least-squares fit, referring to the last digits. $b$ is given in \si{\per\micro\metre}; $a$ (exponential), $a$ (power-law, in \si{\micro\metre}$^{n}$) and $n$ are as indicated. Lower BIC values indicate a better model fit}
\label{tab:fit_comparison}
\begin{resultstabular}{c|cccc|cccc|c}
\hline
& \multicolumn{4}{c|}{\textbf{Exponential}} & \multicolumn{4}{c|}{\textbf{Power-law}} & \\
\cmidrule(lr){2-5} \cmidrule(lr){6-9}
$\boldsymbol{H_{\mathrm{dim}}}$ & $\boldsymbol{a}$ & $\boldsymbol{b}$ & $\boldsymbol{R^2}$ & \textbf{BIC} & $\boldsymbol{a}$ & $\boldsymbol{n}$ & $\boldsymbol{R^2}$ & \textbf{BIC} & \textbf{\makecell{Preferred\\model}} \\ \hline
$H_0$ & 0.700(23) & 0.1283(20) & 0.982 & $-326.9$ & 6.7(6) & 1.62(3) & 0.973 & $-295.0$ & Exponential \\
$H_1$ & 0.357(15) & 0.1255(24) & 0.972 & $-295.6$ & 3.3(3) & 1.59(3) & 0.970 & $-290.2$ & Exponential \\
$H_2$ & 0.564(21) & 0.1473(22) & 0.983 & $-308.4$ & 8.3(9) & 1.90(4) & 0.967 & $-256.9$ & Exponential \\ \hline
\end{resultstabular}
\end{table}

The fitted exponential relationship also provides a direct way to relate the observation size directly to the remaining topological error. For the exponential model, given two subvolume sizes $L_1$ and $L_2 = L_1 +\Delta L$, the ratio of their distances to the reference is:
\begin{equation}
r = \frac{\hat{D}_{\exp}(L_2)}{\hat{D}_{\exp}(L_1)} = \exp(-b \Delta L),
\end{equation}
which can be inverted to give the size increment $\Delta L$ required to achieve a prescribed relative remaining error $r$ in the topological distance:
\begin{equation}
\Delta L = -\frac{\ln r}{b}.
\label{eq:delta_L_exp}
\end{equation}
This expression depends only on the target remaining error ratio $r$ and the decay rate $b$, and is independent of the initial size $L_1$. For example, reducing the size-dependent error by \SI{50}{\percent} ($r=0.5$) requires a size increment of $\Delta L = \ln(2)/b$. Therefore, once the decay rate has been determined, the exponential model provides a practical estimate of the additional observation length required to achieve a desired level of convergence. For a fixed subvolume aspect ratio, this length can subsequently be related to the corresponding observation volume.

Since each homology dimension captures different topological features of the microstructure, a subvolume should be representative with respect to all three simultaneously. The conservative estimate can therefore determined by the slowest-converging dimension, corresponding to the smallest fitted values of $b=0.1255$ (\cref{tab:fit_comparison}). Using this parameter in \cref{eq:delta_L_exp} provides the observation-size increment required to achieve a prescribed reduction in topological error across all topological features. As an illustrative example, consider increasing the shortest subvolume edge from \SI{10}{\micro\metre} to \SI{20}{\micro\metre}. Using the conservative parameters, the exponential model predicts a reduction in topological error of approximately \SI{72}{\percent}. This provides a quantitative criterion for selecting the observation size based on the acceptable topological error.

\begin{figure}[htbp]
    \centering
    \includegraphics[
       width=\linewidth,
       height=0.62\textheight,
       keepaspectratio
   ]{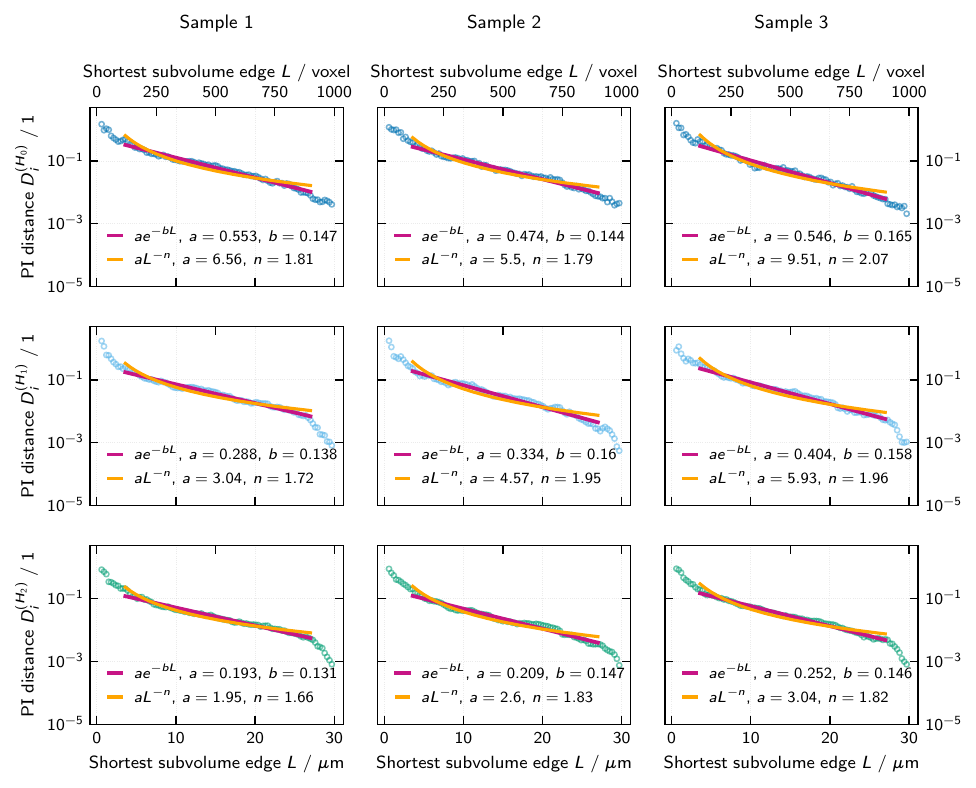}
    \caption{Self-reference PI distance of the three synthetic samples (columns: samples 1--3) as a function of the shortest subvolume edge $L$, for $H_0$ (top row), $H_1$ (middle row) and $H_2$ (bottom row). Circles denote the data, lines the exponential ($\hat{D}_{\mathrm{exp}} = a\,\exp(-bL)$, magenta) and power-law ($\hat{D}_{\mathrm{pow}} = a\,L^{-n}$, orange) fits, shown together with the fitted parameters. The vertical axis is logarithmic. The top axis gives $L$ in voxels. Fit parameters are listed in \cref{tab:fit_comparison_synthetic}}
    \label{fig:syntetic_selfref}
\end{figure}

To assess whether the observed convergence behavior is also obtained for structures with controlled statistical characteristics, the same analysis was performed for three independent synthetic realizations of a stochastic two-phase microstructure. For all three realizations, the PI distance decreased with increasing observation volume (\cref{fig:syntetic_selfref}), and under the self-reference scheme the exponential model was preferred over the power-law model for all three homology dimensions, i.e., in all nine analyzed cases (\cref{tab:fit_comparison_synthetic}). Thus, both the decrease in PI distance and the preference for the exponential model observed for the experimental microstructure were reproduced for independently generated microstructures. This exponential model described the data well, with $R^2$ between 0.975 and 0.992 (\cref{tab:fit_comparison_synthetic}), and was preferred over the power-law model by a wide margin ($\Delta\mathrm{BIC}$ from -26 to -145). The decay rate $b$ varied only moderately across realizations and homology dimensions, from \SIrange{0.197}{0.248}{\per\micro\metre}, with a mean of \SI{0.223}{\per\micro\metre} and a standard deviation of \SI{0.016}{\per\micro\metre} (\SI{7}{\percent} of the mean), indicating a similar convergence rate for microstructures generated with the same statistical parameters. With only three realizations, it cannot be determined whether this variation reflects a genuine dependence on the particular spatial arrangement or ordinary sample-to-sample fluctuation. Nevertheless, it suggests that the convergence rate should be estimated for the specific microstructure of interest rather than assumed from its nominal statistical parameters.
More importantly, the fitted $b$ should not be interpreted as a universal constant. It differs between the experimental microstructure (\SIrange{0.126}{0.147}{\per\micro\metre}) and the synthetic realizations (\SIrange{0.197}{0.248}{\per\micro\metre}), and also varies between individual synthetic realizations. It also depends on the choice of reference: the same subvolumes give different values of $b$ when compared against the PI of the whole sample (self-reference) than when compared against the PI of a different sample (cross-reference).

 \begin{table}[htbp]
   \centering
   \footnotesize
   \setlength{\tabcolsep}{4pt}
\caption{Parameters, $R^2$ and BIC values of the exponential ($\hat{D}_{\mathrm{exp}} = a\,\exp(-bL)$) and power-law ($\hat{D}_{\mathrm{pow}} = a\,L^{-n}$) fits applied to the PI distance curves of the synthetic microstructures (self-reference), for individual homology dimensions $H_{\mathrm{dim}}$ and synthetic samples 1--3. Fits, $R^2$ and BIC computed in log-space. Values in parentheses are standard errors of the least-squares fit, referring to the last digits. $b$ is given in \si{\per\micro\metre}; $a$ (exponential), $a$ (power-law, in \si{\micro\metre}$^{n}$) and $n$ are as indicated. Lower BIC values indicate a better model fit}
\label{tab:fit_comparison_synthetic}
\begin{resultstabular}{c|cccc|cccc|c}
\hline
& \multicolumn{4}{c|}{\textbf{Exponential}} & \multicolumn{4}{c|}{\textbf{Power-law}} & \\
\cmidrule(lr){2-5} \cmidrule(lr){6-9}
$\boldsymbol{H_{\mathrm{dim}}}$ & $\boldsymbol{a}$ & $\boldsymbol{b}$ & $\boldsymbol{R^2}$ & \textbf{BIC} & $\boldsymbol{a}$ & $\boldsymbol{n}$ & $\boldsymbol{R^2}$ & \textbf{BIC} & \textbf{\makecell{Preferred\\model}} \\ \hline
\multicolumn{10}{c}{\textbf{Sample 1}} \\ \hline
$H_0$ & 0.553(19) & 0.2199(31) & 0.985 & $-321.5$ & 3.1(4) & 1.81(6) & 0.934 & $-204.0$ & Exponential \\
$H_1$ & 0.288(10) & 0.2071(31) & 0.983 & $-320.5$ & 1.5(2) & 1.72(5) & 0.944 & $-226.6$ & Exponential \\
$H_2$ & 0.193(7) & 0.1968(32) & 0.980 & $-314.9$ & 0.99(7) & 1.66(3) & 0.972 & $-289.2$ & Exponential \\ \hline
\multicolumn{10}{c}{\textbf{Sample 2}} \\ \hline
$H_0$ & 0.474(16) & 0.2166(30) & 0.985 & $-324.7$ & 2.7(3) & 1.79(5) & 0.940 & $-214.4$ & Exponential \\
$H_1$ & 0.334(15) & 0.2399(40) & 0.979 & $-280.7$ & 2.1(4) & 1.95(7) & 0.899 & $-156.5$ & Exponential \\
$H_2$ & 0.209(9) & 0.2201(40) & 0.975 & $-280.7$ & 1.2(1) & 1.83(5) & 0.942 & $-213.7$ & Exponential \\ \hline
\multicolumn{10}{c}{\textbf{Sample 3}} \\ \hline
$H_0$ & 0.546(22) & 0.2482(37) & 0.983 & $-294.2$ & 4.1(5) & 2.07(5) & 0.955 & $-215.8$ & Exponential \\
$H_1$ & 0.404(13) & 0.2370(30) & 0.988 & $-327.6$ & 2.7(3) & 1.96(5) & 0.945 & $-206.6$ & Exponential \\
$H_2$ & 0.252(6) & 0.2196(23) & 0.992 & $-370.4$ & 1.5(2) & 1.82(5) & 0.949 & $-225.1$ & Exponential \\ \hline
\end{resultstabular}
\end{table}

\begin{figure}[htbp]
    \centering
    \includegraphics[
       width=0.85\linewidth,
       height=0.62\textheight,
       keepaspectratio
   ]{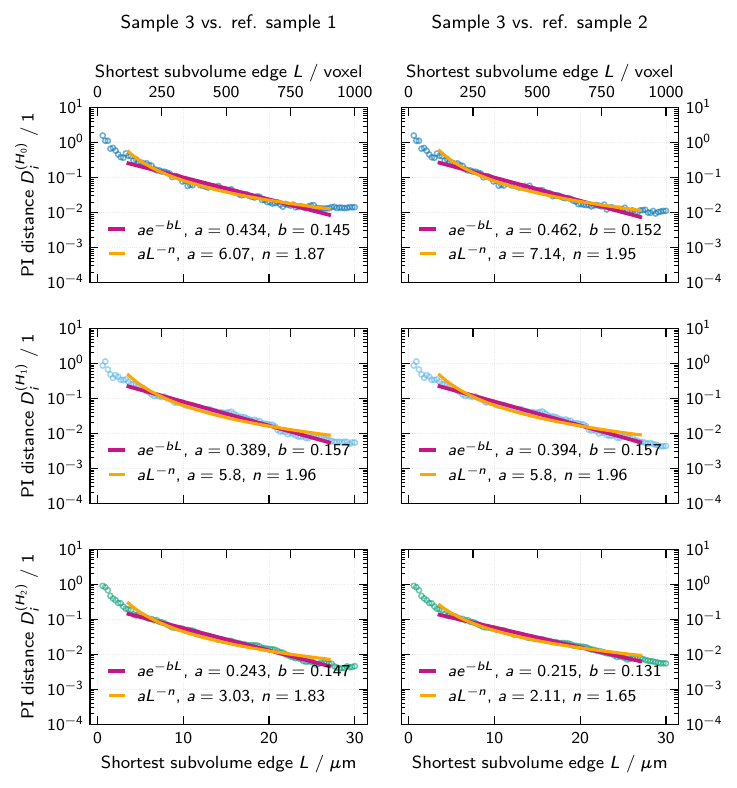}
    \caption{PI distance of synthetic sample 3 evaluated against the largest subvolume of reference sample 1 (left column) and reference sample 2 (right column) as a function of the shortest subvolume edge $L$, for $H_0$ (top row), $H_1$ (middle row) and $H_2$ (bottom row). Circles denote the data, lines the exponential ($\hat{D}_{\mathrm{exp}} = a\,\exp(-bL)$, magenta) and power-law ($\hat{D}_{\mathrm{pow}} = a\,L^{-n}$, orange) fits, shown together with the fitted parameters. The vertical axis is logarithmic. The top axis gives $L$ in voxels. Fit parameters for all sample--reference pairs are listed in \cref{tab:fit_comparison_crossref}.}
    \label{fig:crossref_sample3}
\end{figure}

The nested-sampling framework raises a question of whether the observed preference for the exponential model reflects an intrinsic property of the microstructure's topology or is, at least partially, a consequence of how subvolumes are constructed within a single nested reference. To examine this, the analysis for the synthetic realizations was repeated using cross-reference distances, in which each realization was compared not to its own largest subvolume, but to the largest subvolume of a different, statistically equivalent realization (\cref{fig:crossref_sample3}). The exponential decay model was still preferred in eight of the twelve analyzed cases (\cref{tab:fit_comparison_crossref}), including all four cases for $H_1$. For $H_0$ and $H_2$, however, the power-law model was preferred in two out of four cross-reference pairs each. The exponential model therefore does not describe the convergence equally well in every reference configuration: the power-law model can outperform it when the reference is an independent realization.

 \begin{table}[htbp]
   \centering
   \footnotesize
   \setlength{\tabcolsep}{4pt}
\caption{Parameters, $R^2$ and BIC values of the exponential ($\hat{D}_{\mathrm{exp}} = a\,\exp(-bL)$) and power-law ($\hat{D}_{\mathrm{pow}} = a\,L^{-n}$) fits applied to the cross-reference PI distance curves of the synthetic microstructures, for individual homology dimensions $H_{\mathrm{dim}}$. Each block gives the PI distance of the indicated sample evaluated against the indicated reference sample. Fits, $R^2$ and BIC computed in log-space. Values in parentheses are standard errors of the least-squares fit, referring to the last digits. $b$ is given in \si{\per\micro\metre}; $a$ (exponential), $a$ (power-law, in \si{\micro\metre}$^{n}$) and $n$ are as indicated. Lower BIC values indicate a better model fit}
\label{tab:fit_comparison_crossref}
\begin{resultstabular}{c|cccc|cccc|c}
\hline
& \multicolumn{4}{c|}{\textbf{Exponential}} & \multicolumn{4}{c|}{\textbf{Power-law}} & \\
\cmidrule(lr){2-5} \cmidrule(lr){6-9}
$\boldsymbol{H_{\mathrm{dim}}}$ & $\boldsymbol{a}$ & $\boldsymbol{b}$ & $\boldsymbol{R^2}$ & \textbf{BIC} & $\boldsymbol{a}$ & $\boldsymbol{n}$ & $\boldsymbol{R^2}$ & \textbf{BIC} & \textbf{\makecell{Preferred\\model}} \\ \hline
\multicolumn{10}{c}{\textbf{Sample 1 vs.\ ref.\ sample 2}} \\ \hline
$H_0$ & 0.490(15) & 0.2038(28) & 0.986 & $-339.1$ & 2.6(2) & 1.70(4) & 0.961 & $-257.9$ & Exponential \\
$H_1$ & 0.278(9) & 0.1993(30) & 0.983 & $-327.5$ & 1.4(1) & 1.66(4) & 0.951 & $-243.2$ & Exponential \\
$H_2$ & 0.168(7) & 0.1744(39) & 0.963 & $-284.1$ & 0.76(3) & 1.50(2) & 0.989 & $-382.5$ & Power-law \\ \hline
\multicolumn{10}{c}{\textbf{Sample 2 vs.\ ref.\ sample 1}} \\ \hline
$H_0$ & 0.413(14) & 0.1918(29) & 0.982 & $-329.4$ & 2.0(2) & 1.61(3) & 0.968 & $-282.2$ & Exponential \\
$H_1$ & 0.308(10) & 0.2310(30) & 0.987 & $-327.7$ & 1.9(3) & 1.89(6) & 0.923 & $-184.8$ & Exponential \\
$H_2$ & 0.180(8) & 0.2000(40) & 0.971 & $-281.9$ & 0.98(6) & 1.70(3) & 0.981 & $-318.2$ & Power-law \\ \hline
\multicolumn{10}{c}{\textbf{Sample 3 vs.\ ref.\ sample 1}} \\ \hline
$H_0$ & 0.43(3) & 0.217(5) & 0.957 & $-237.5$ & 2.85(17) & 1.87(3) & 0.984 & $-318.2$ & Power-law \\
$H_1$ & 0.390(14) & 0.2358(32) & 0.986 & $-317.6$ & 2.6(3) & 1.96(5) & 0.953 & $-219.9$ & Exponential \\
$H_2$ & 0.243(6) & 0.2202(24) & 0.991 & $-362.9$ & 1.4(1) & 1.83(4) & 0.959 & $-242.5$ & Exponential \\ \hline
\multicolumn{10}{c}{\textbf{Sample 3 vs.\ ref.\ sample 2}} \\ \hline
$H_0$ & 0.46(2) & 0.229(5) & 0.968 & $-254.8$ & 3.2(2) & 1.95(3) & 0.982 & $-297.8$ & Power-law \\
$H_1$ & 0.394(13) & 0.2359(29) & 0.988 & $-329.3$ & 2.6(3) & 1.96(5) & 0.950 & $-215.6$ & Exponential \\
$H_2$ & 0.216(6) & 0.1958(26) & 0.987 & $-351.5$ & 1.08(8) & 1.65(3) & 0.972 & $-290.0$ & Exponential \\ \hline
\end{resultstabular}
\end{table}

For practical tomographic analysis, an independent statistically equivalent reference volume is generally unavailable, as the reconstructions are rarely perfectly homogeneous. The largest accessible subvolume therefore provides a practical reference for quantifying the change in topological representation with increasing observation size. Within this self-reference framework, the exponential model consistently describes the observed convergence for the experimental microstructure and all three synthetic realizations. The fitted parameters $a$ and $b$ should consequently be regarded as effective, sample-specific descriptors of topological convergence rather than universal material constants. Within this scope, they provide a quantitative relation between observation volume and topological representativeness error for a given reconstructed microstructure and sampling procedure.
\section{Conclusion}
 \label{sec:conlusion}
In this work, we introduced a topology-based framework for representative volume element determination in porous solid oxide fuel cell anodes using descriptors derived from persistent homology. By analyzing hierarchical subvolumes extracted from one of the largest reported three-dimensional PFIB-SEM datasets of a Ni-YSZ anode, we evaluated structural convergence as a function of observation scale. Topological similarity was quantified by measuring the $L^1$ distance between the normalized persistence image vectors of each subvolume and that of the largest available reference volume.

The topological distance decreases monotonically with increasing subvolume size, demonstrating progressive convergence toward the topology of the reference volume. This continuous convergence enables representative volume element estimation to be formulated in terms of the remaining topological error rather than as a binary representative/non-representative decision. The proposed framework therefore quantifies \textit{how representative} a given observation volume is relative to a large reference volume. Because this convergence is accurately described by an exponential decay model ($D_i = a\,e^{-bL}$), with high agreement across all homology dimensions ($R^2 \in [0.972, 0.983]$), it allows the minimum subvolume size required to achieve any prescribed representativeness error to be determined analytically. For instance, using the conservative decay rate ($b = \SI{0.1255}{\per\micro\metre}$ for $H_1$), expanding the shortest subvolume edge from \SI{10}{\micro\metre} to \SI{20}{\micro\metre} yields an approximate \SI{72}{\percent} reduction in topological error. The large volume of our reconstruction also has an important practical implication: conventional FIB-SEM datasets appear to lie close to the minimum size required even for the convergence of classical geometric descriptors, suggesting that previously reported RVEs may have been estimated from datasets where the accessible volume was insufficient to fully assess convergence.

The synthetic validation reproduced the decrease in PI distance and the preference for the exponential model in all nine self-reference cases across three independent realizations. However, the fitted decay rate $b$ varied between realizations, and in the cross-reference analysis the power-law model was preferred in four of twelve cases. The exponential relationship should therefore be regarded as an empirical description of convergence within the investigated observation-size range and sampling configuration, rather than as a universal scaling law. The fitted parameters $a$ and $b$ are sample-specific descriptors that provide a quantitative relation between observation size and topological representativeness error, and are expected to vary with material, microstructure, and processing conditions.

Because the analyzed reconstruction exhibits microstructural gradients, extending the current methodology toward the quantitative characterization of microstructural heterogeneity is a natural direction for future work. The present work demonstrates the proposed framework using a single large PFIB-SEM reconstruction of a Ni-YSZ anode. The framework, however, is formulated independently of the specific material system and can, in principle, be applied to other porous and multiphase microstructures. More broadly, this study demonstrates that representativeness can be treated as a continuous measure of structural similarity that evolves with the observation scale. By providing a quantitative link between observation volume and representativeness error, the proposed topology-based framework offers a new perspective for RVE determination in porous and multiphase materials exhibiting scale-dependent behavior.
\section*{Conflict of Interest}
   The authors declare no competing interests.

\section*{Acknowledgments}
 This work was supported by the Polish National Agency for Academic Exchange (NAWA), within the Strategic Partnerships Programme, Project No. BPI/PST/2024/1/00046 (project title: Research and Educational Cooperation in the field of Energy and Environmental Engineering: AGH's strategic partnership with Japanese universities). The authors acknowledge the support of the Polish Ministry of Science and Higher Education program “Excellence Initiative—Research University” for the AGH University of Krakow, Poland, project no. UB-696000-00001-Z01. We gratefully acknowledge Polish high-performance computing infrastructure PLGrid (HPC Center: ACK Cyfronet AGH) for providing computer facilities and support within computational grant no. PLG/2026/019600

\section*{Declaration of Generative AI and AI-assisted technologies in the writing process}
    During the preparation of this work, the authors used Gemini AI and OpenAI ChatGPT to refine English writing and improve language clarity. After using this tool, the authors reviewed and edited the content as needed and took full responsibility for the content of the publication.
\section*{CRediT}
    \textbf{Aleksandra Kukiełka}: conceptualization, methodology,  software (subvolumes extraction, persistence images, topological distance analysis, synthetic microstructures generation, model fitting), formal analysis, investigation, writing --- original draft, writing --- review \& editing, visualization, funding acquisition.
    \textbf{Szymon Buchaniec}: conceptualization, supervision, writing --- review \& editing, methodology, software (persistence images). \textbf{Maksym Szemer}: software (Cubicle setup, persistence diagrams: birth-death pairs, processing automation), writing --- original draft
    \textbf{Katarzyna Berent}:  investigation (PFIB-SEM), resources (PFIB-SEM), writing --- original draft 
     \textbf{Grzegorz Brus}: conceptualization, supervision, writing -- review \& editing, writing --- original draft, methodology, Investigation (phase segmentation and reconstruction), formal analysis (Microstructure segmentation and reconstruction), supervision, project administration, funding acquisition, resources (AVIZO Thermofisher Scientific).
    
\section{Appendix}
\subsection{Homogeneity assessment}
\label{subsec:homogeneity}

\begin{figure}[htbp]
    \centering
    \begin{subfigure}[t]{0.48\textwidth}
        \centering
         \includegraphics[width=\linewidth]{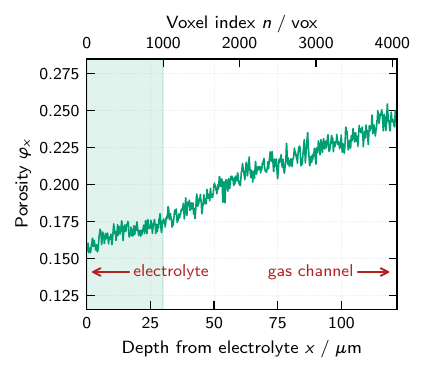}
        \caption{Porosity profile along the $x$-axis, corresponding to the direction from the electrolyte toward the gas channel}
        \label{fig:porosity_x}
    \end{subfigure}
    \hfill
    \begin{subfigure}[t]{0.48\textwidth}
        \centering
        \includegraphics[width=\linewidth]{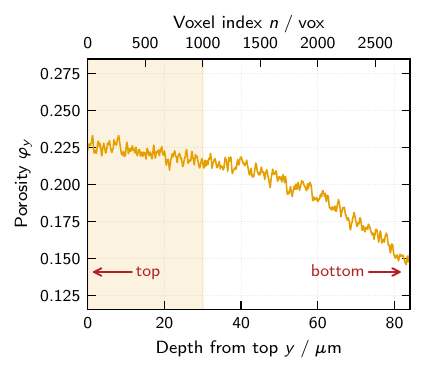}
        \caption{Porosity profile along the $y$-axis of the full reconstructed volume, corresponding to the direction from the top to bottom of the sample}
        \label{fig:porosity_y}
    \end{subfigure}
     \par\medskip

    \begin{subfigure}[t]{0.48\textwidth}
        \centering
        \includegraphics[width=\linewidth]{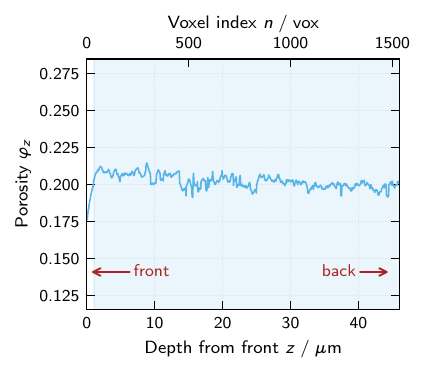}
        \caption{Porosity profile along the $z$-axis}
        \label{fig:porosity_z}
    \end{subfigure}

    \caption{Slice-by-slice porosity profiles along the three principal axes of the full reconstructed volume. Shaded areas indicate the subvolume selected for further analysis}
    \label{fig:porosity_profiles}
\end{figure}

Slice-by-slice porosity profiles were evaluated along the three principal axes of the reconstructed volume in order to identify the largest quasi-homogeneous region suitable for subsequent RVE analysis. As shown in \cref{fig:porosity_profiles}, the reconstructed dataset exhibits systematic porosity gradients along all three directions, indicating that no globally homogeneous domain is present.

A pronounced monotonic porosity gradient was observed along the $x$-axis (\cref{fig:porosity_x}), corresponding to the direction from the electrolyte toward the gas channel. Such behavior is consistent with the functional grading commonly reported for SOFC anodes, where the pore structure varies across the electrode thickness as a consequence of fabrication.

Along the $y$-axis (\cref{fig:porosity_y}), a systematic decrease in porosity was also detected. This trend is attributed to the imaging process rather than to the intrinsic microstructure. During large-volume PFIB-SEM serial sectioning, the substantial amount of removed material leads to progressive debris accumulation toward the bottom of the sample, partially obstructing pores in the lower slices and thereby reducing the apparent local porosity. The magnitude of this artifact is expected to increase with reconstructed sample height and therefore represents an important practical limitation of large-volume PFIB-SEM tomography.

In contrast, the porosity profile along the $z$-axis remained approximately stationary over most of the reconstructed volume (\cref{fig:porosity_z}), exhibiting fluctuations within approximately $\pm1\%$ around the mean. Only the outermost region ($z<36$ voxels) showed noticeable deviations caused by surface preparation artifacts and was therefore excluded from further analysis.

Based on these observations, a subvolume spanning $x \in [0, 1000] $, $y \in [0, 1000]$, and $z \in [36, 1536]$  voxels was selected for the subsequent analyses, corresponding to physical dimensions of \SI{30}{\micro\metre} $\times$ \SI{30}{\micro\metre} $\times$ \SI{45}{\micro\metre}
(\qtyproduct[product-units = single] {1000 x 1000 x 1500} {voxels}). This region provides the largest available domain with the lowest porosity gradients along all three directions, thereby minimizing systematic bias while maximizing the observation volume. Despite this cropping, the selected region remains more than twice the volume of the largest three-dimensional SOFC anode reconstruction reported using conventional FIB-SEM~\cite{Kanno.2011}, while preserving comparable nanoscale resolution. Nevertheless, because neither the functional gradient along $z$ nor the acquisition-induced gradient along $y$ could be completely eliminated, the selected domain should be regarded as quasi-homogeneous rather than perfectly homogeneous. Consequently, all subsequent analyses were performed within the largest quasi-homogeneous region rather than over the entire reconstructed volume.

\clearpage
\bibliography{bibliography}

@article{Kusano.2017,
  author  = {Genki Kusano and Kenji Fukumizu and Yasuaki Hiraoka},
  title   = {Kernel Method for Persistence Diagrams via Kernel Embedding and Weight Factor},
  journal = {Journal of Machine Learning Research},
  year    = {2018},
  volume  = {18},
  number  = {189},
  pages   = {1--41},
}

@article{Adams.2017,
  title={Persistence images: A stable vector representation of persistent homology},
  author={Adams, Henry and Emerson, Tegan and Kirby, Michael and Neville, Rachel and Peterson, Chris and Shipman, Patrick and Chepushtanova, Sofya and Hanson, Eric and Motta, Francis and Ziegelmeier, Lori},
  journal={Journal of Machine Learning Research},
  volume={18},
  number={8},
  pages={1--35},
  year={2017}
}

@article{Ruse.2024, 
  year     = {2024}, 
  title    = {Quantifying Microstructure Features for High-Performance Solid Oxide Cells}, 
  author   = {Ruse, Cristina Mariana and Hume, Lily Ann and Wang, Yudong and Pesacreta, Thomas C. and Zhou, Xiao-Dong}, 
  journal  = {Materials}, 
  issn     = {1996-1944}, 
  doi      = {10.3390/ma17112622}, 
  pages    = {2622}, 
  number   = {11}, 
  volume   = {17}
}

@article{Joos.2012, 
  year     = {2012}, 
  title    = {Representative volume element size for accurate solid oxide fuel cell cathode reconstructions from focused ion beam tomography data}, 
  author   = {Joos, Jochen and Ender, Moses and Carraro, Thomas and Weber, André and Ivers-Tiffée, Ellen}, 
  journal  = {Electrochimica Acta}, 
  issn     = {0013-4686}, 
  doi      = {10.1016/j.electacta.2012.04.133},
  pages    = {268--276}, 
  volume   = {82}
}

@article{Epting.2017, 
  year     = {2017}, 
  title    = {Quantifying intermediate‐frequency heterogeneities of {SOFC} electrodes using X‐ray computed tomography}, 
  author   = {Epting, William K. and Mansley, Zachary and Menasche, David B. and Kenesei, Peter and Suter, Robert M. and Gerdes, Kirk and Litster, Shawn and Salvador, Paul A.}, 
  journal  = {Journal of the American Ceramic Society}, 
  issn     = {0002-7820}, 
  doi      = {10.1111/jace.14775},
  pages    = {2232--2242}, 
  number   = {5}, 
  volume   = {100}
}

@article{Joos.2011, 
  year     = {2011}, 
  title    = {Detailed Microstructure Analysis and 3D Simulations of Porous Electrodes}, 
  author   = {Joos, Jochen and Carraro, Thomas and Ender, Moses and Rüger, Bernd and Weber, André and Ivers-Tiffée, Ellen}, 
  journal  = {Electrochemical Society Transactions}, 
  issn     = {1938-5862}, 
  doi      = {10.1149/1.3570232},
  pages    = {2357--2368}, 
  number   = {1}, 
  volume   = {35}
}

@article{Kanit.2003, 
  year     = {2003}, 
  title    = {Determination of the size of the representative volume element for random composites: statistical and numerical approach}, 
  author   = {Kanit, T. and Forest, S. and Galliet, I. and Mounoury, V. and Jeulin, D.}, 
  journal  = {International Journal of Solids and Structures}, 
  issn     = {0020-7683}, 
  doi      = {10.1016/s0020-7683(03)00143-4},
  pages    = {3647--3679}, 
  number   = {13-14}, 
  volume   = {40}
}

@article{Laurencin.2012, 
  year     = {2012}, 
  title    = {Characterisation of Solid Oxide Fuel Cell Ni–8YSZ substrate by synchrotron X-ray nano-tomography: from 3D reconstruction to microstructure quantification}, 
  author   = {Laurencin, J. and Quey, R. and Delette, G. and Suhonen, H. and Cloetens, P. and Bleuet, P.}, 
  journal  = {Journal of Power Sources}, 
  issn     = {0378-7753}, 
  doi      = {10.1016/j.jpowsour.2011.09.105},
  pages    = {182--189}, 
  volume   = {198}
}

@article{Harris.2015, 
  year     = {2015}, 
  title    = {Determining the representative volume element size for three-dimensional microstructural material characterization. Part 2: Application to experimental data}, 
  author   = {Harris, William M. and Chiu, Wilson K.S.}, 
  journal  = {Journal of Power Sources}, 
  issn     = {0378-7753}, 
  doi      = {10.1016/j.jpowsour.2015.02.052},
  pages    = {622--629}, 
  volume   = {282}
}

@article{Mahbub.2021, 
  year     = {2021}, 
  title    = {Quantifying morphological variability and operating evolution in {SOFC} anode microstructures}, 
  author   = {Mahbub, Rubayyat and Hsu, Tim and Epting, William K. and Nolan, Gillian and Lei, Yinkai and Nuhfer, Noel T. and Doane, Randell B. and Abernathy, Harry W. and Hackett, Gregory A. and Litster, Shawn and Rollett, Anthony D. and Salvador, Paul A.}, 
  journal  = {Journal of Power Sources}, 
  issn     = {0378-7753}, 
  doi      = {10.1016/j.jpowsour.2021.229846}, 
  pages    = {229846}, 
  volume   = {498}
}

@article{Zekri.2017, 
  year     = {2017}, 
  title    = {Microstructure degradation of Ni/{CGO} anodes for solid oxide fuel cells after long operation time using 3D reconstructions by {FIB} tomography}, 
  author   = {Zekri, Atef and Knipper, Martin and Parisi, Jürgen and Plaggenborg, Thorsten}, 
  journal  = {Physical Chemistry Chemical Physics}, 
  issn     = {1463-9076}, 
  doi      = {10.1039/c7cp02186k}, 
  pmid     = {28504292}, 
  pages    = {13767--13777}, 
  number   = {21}, 
  volume   = {19}
}

@inproceedings{Wagner.2023.cubicle,
  title={Slice, Simplify and Stitch: Topology-Preserving Simplification Scheme for Massive Voxel Data},
  author={Wagner, Hubert},
  booktitle={39th International Symposium on Computational Geometry (SoCG 2023)},
  year={2023},
  organization={Schloss Dagstuhl-Leibniz-Zentrum f{\"u}r Informatik}
}

@article{Yamatoko.2025, 
  year     = {2025}, 
  title    = {Advanced 3D Microstructure Generation of Solid Oxide Cell Electrodes Using Conditional Generative Adversarial Network and Validation Using Nonintuitive Topological Characteristics}, 
  author   = {Yamatoko, Sojiro and Kishimoto, Masashi and Buchaniec, Szymon and Guo, Yuting and Iwai, Hiroshi}, 
  journal  = {Advanced Intelligent Discovery}, 
  issn     = {2943-9981}, 
  doi      = {10.1002/aidi.202500010}, 
  number   = {2}, 
  volume   = {1}
}

@article{Szemer.2025,
  year     = {2025}, 
  title    = {Topology-informed machine learning for efficient prediction of solid oxide fuel cell electrode polarization}, 
  author   = {Szemer, Maksym and Buchaniec, Szymon and Prokop, Tomasz and Brus, Grzegorz}, 
  journal  = {Energy and AI}, 
  issn     = {2666-5468}, 
  doi      = {10.1016/j.egyai.2025.100495}, 
  eprint   = {2410.05307}, 
  pages    = {100495}, 
  volume   = {20}
}

@article{Kanno.2011, 
  year     = {2011}, 
  title    = {Evaluation of SOFC anode polarization simulation using three-dimensional microstructures reconstructed by FIB tomography}, 
  author   = {Kanno, Daisuke and Shikazono, Naoki and Takagi, Norikazu and Matsuzaki, Katsuhisa and Kasagi, Nobuhide}, 
  journal  = {Electrochimica Acta}, 
  issn     = {0013-4686}, 
  doi      = {10.1016/j.electacta.2011.02.010}, 
  pages    = {4015--4021}, 
  number   = {11}, 
  volume   = {56}
}

@article{Yoshida.2024,
    author = {Yoshida, Ryuji and Mizuta, Yasutoshi and Kato, Takeharu and Kimura, Teiichi},
    title = {Comparison of large-volume 3D reconstruction using plasma FIB-SEM and X-ray CT},
    journal = {Microscopy},
    volume = {73},
    number = {4},
    pages = {354-357},
    year = {2024},
    month = {01},
    issn = {2050-5701},
    doi = {10.1093/jmicro/dfae003},
    url = {https://doi.org/10.1093/jmicro/dfae003},
    eprint = {https://academic.oup.com/jmicro/article-pdf/73/4/354/59594068/dfae003.pdf},
}

@article{Wilson.2011, 
  year     = {2011}, 
  title    = {Linking the microstructure, performance and durability of Ni-yttria-stabilized zirconia solid oxide fuel cell anodes using three-dimensional focused ion beam–scanning electron microscopy imaging}, 
  author   = {Wilson, James R. and Cronin, J. Scott and Barnett, Scott A.}, 
  journal  = {Scripta Materialia}, 
  issn     = {1359-6462}, 
  doi      = {10.1016/j.scriptamat.2010.09.025}, 
  pages    = {67--72}, 
  number   = {2}, 
  volume   = {65}
}

@article{Obayashi.2022.HomCloud, 
  year     = {2022}, 
  title    = {Persistent Homology Analysis for Materials Research and Persistent Homology Software: {HomCloud}}, 
  author   = {Obayashi, Ippei and Nakamura, Takenobu and Hiraoka, Yasuaki}, 
  journal  = {Journal of the Physical Society of Japan}, 
  issn     = {0031-9015}, 
  doi      = {10.7566/jpsj.91.091013}, 
  eprint   = {2112.03610},
  pages    = {091013}, 
  number   = {9}, 
  volume   = {91}
}

@article{Edelsbrunner.2008,
  title={Persistent homology-a survey},
  author={Edelsbrunner, Herbert and Harer, John and others},
  journal={Contemporary mathematics},
  volume={453},
  number={26},
  pages={257--282},
  year={2008},
  publisher={Providence, RI: American Mathematical Society}
}

@article{chen.2017, 
  year     = {2017}, 
  title    = {Convolutional neural networks for automated annotation of cellular cryo-electron tomograms}, 
  author   = {Chen, Muyuan and Dai, Wei and Sun, Stella Y and Jonasch, Darius and He, Cynthia Y and Schmid, Michael F and Chiu, Wah and Ludtke, Steven J}, 
  journal  = {Nature Methods}, 
  issn     = {1548-7091}, 
  doi      = {10.1038/nmeth.4405}, 
  pmid     = {28846087}, 
  pmcid    = {{PMC}5623144}, 
  eprint   = {1701.05567}, 
  pages    = {983--985}, 
  number   = {10}, 
  volume   = {14}
}

@article{skimage.2014,
  author    = {van der Walt, Stefan and others},
  title     = {scikit-image: image processing in {Python}},
  journal   = {PeerJ},
  volume    = {2},
  pages     = {e453},
  year      = {2014},
  doi       = {10.7717/peerj.453}
}

@article{Simonyan.2014, 
  year     = {2014}, 
  title    = {Very Deep Convolutional Networks for Large-Scale Image Recognition}, 
  author   = {Simonyan, Karen and Zisserman, Andrew}, 
  journal  = {{arXiv}}, 
  doi      = {10.48550/arxiv.1409.1556}, 
  eprint   = {1409.1556}
}

@article{Vafaeenezhad.2022, 
  year     = {2022}, 
  title    = {Microstructure and long-term stability of Ni–{YSZ} anode supported fuel cells: a review}, 
  author   = {Vafaeenezhad, Sajad and Hanifi, Amir Reza and Laguna-Bercero, Miguel A and Etsell, Thomas H and Sarkar, Partha}, 
  journal  = {Materials Futures}, 
  doi      = {10.1088/2752-5724/ac88e7},  
  pages    = {042101}, 
  number   = {4}, 
  volume   = {1}
}

@article{Hsu.2018,
    title={Mesoscale characterization of local property distributions in heterogeneous electrodes},
    volume={386},
    ISSN={0378-7753},
    DOI={10.1016/j.jpowsour.2018.03.025}, journal={Journal of Power Sources}, author={Hsu, Tim and Epting, William K. and Mahbub, Rubayyat and Nuhfer, Noel T. and Bhattacharya, Sudip and Lei, Yinkai and Miller, Herbert M. and Ohodnicki, Paul R. and Gerdes, Kirk R. and Abernathy, Harry W. and Hackett, Gregory A. and Rollett, Anthony D. and Graef, Marc De and Litster, Shawn and Salvador, Paul A.},
    year={2018},
    pages={1–9}
}

@article{Pawowski2023, 
  year    = {2023}, 
  title   = {Microstructure evolution of Solid Oxide Fuel Cell anodes characterized by persistent homology}, 
  author  = {Pawłowski, Piotr and Buchaniec, Szymon and Prokop, Tomasz and Iwai, Hiroshi and Brus, Grzegorz}, 
  journal = {Energy and {AI}}, 
  issn    = {2666-5468}, 
  doi     = {10.1016/j.egyai.2023.100256}, 
  eprint  = {2212.01576}, 
  pages   = {100256}, 
  volume  = {14}
}

@article{Brus.2023,
    title = {Overcoming a recent impasse in the application of artificial neural networks as solid oxide fuel cells simulator with computational topology},
    journal = {Energy and AI},
    volume = {14},
    pages = {100291},
    year = {2023},
    issn = {2666-5468},
    doi = {https://doi.org/10.1016/j.egyai.2023.100291},
    author = {Grzegorz Brus}
}

@article{Sciazko.2023,
    title = {3D microstructures of solid oxide fuel cell Ni-YSZ anodes with carbon deposition},
    journal = {Chemical Engineering Journal},
    volume = {460},
    pages = {141680},
    year = {2023},
    issn = {1385-8947},
    doi = {https://doi.org/10.1016/j.cej.2023.141680},
    author = {Anna Sciazko and Yosuke Komatsu and Akiko Nakamura and Zhufeng Ouyang and Toru Hara and Naoki Shikazono}
}

@book{DeyWang.2022,
    place={Cambridge}, 
    title={Computational Topology for Data Analysis}, publisher={Cambridge University Press}, 
    author={Dey, Tamal Krishna and Wang, Yusu},
    year={2022}
}

@ARTICLE{Zhang.2023,
	author = {Zhang, Xiaoqiang and Yang, Danan and Xu, Min and Naden, Aaron and Espinoza-Andaluz, Mayken and Li, Tingshuai and Irvine, John T. S. and Andersson, Martin},
	title = {Analysing Tortuosity for Solid Oxide Fuel Cell Anode Material: Experiments and Modeling},
	year = {2023},
	journal = {Journal of the Electrochemical Society},
	volume = {170},
	number = {9},
	doi = {10.1149/1945-7111/acf884},
	type = {Article}
}

@ARTICLE{Hoffrogge.2023,
	author = {Hoffrogge, Paul W. and Schneider, Daniel and Wankmüller, Florian and Meffert, Matthias and Gerthsen, Dagmar and Weber, André and Nestler, Britta and Wieler, Matthias},
	title = {Performance estimation by multiphase-field simulations and transmission-line modeling of nickel coarsening in FIB-SEM reconstructed Ni-YSZ SOFC anodes I: Influence of wetting angle},
	year = {2023},
	journal = {Journal of Power Sources},
	volume = {570},
	doi = {10.1016/j.jpowsour.2023.233031},
	type = {Article}
}

@article{Herring.2018,
author = {Herring, A. L. and Robins, V. and Sheppard, A. P.},
title = {Topological Persistence for Relating Microstructure and Capillary Fluid Trapping in Sandstones},
journal = {Water Resources Research},
volume = {55},
number = {1},
pages = {555-573},
doi = {https://doi.org/10.1029/2018WR022780},
year = {2019}
}

@article{Wang.2020,
author = {Wang, Zhi-Lei and Ogawa, Toshio and Adachi, Yoshitaka},
title = {Property Predictions for Dual-Phase Steels Using Persistent Homology and Machine Learning},
journal = {Advanced Theory and Simulations},
volume = {3},
number = {3},
pages = {1900227},
doi = {https://doi.org/10.1002/adts.201900227},
year = {2020}
}

@article{Pritchard.2023,
  author    = {Pritchard, Ysanne and Sharma, Aikta and Clarkin, Claire and Ogden, Helen and Mahajan, Sumeet and S{\'a}nchez-Garc{\'i}a, Rub{\'e}n J.},
  title     = {Persistent homology analysis distinguishes pathological bone microstructure in non-linear microscopy images},
  journal   = {Scientific Reports},
  year      = {2023},
  volume    = {13},
  number    = {1},
  pages     = {2522},
  issn      = {2045-2322},
  doi       = {10.1038/s41598-023-28985-3}
}

@article{Gao.2024,
title = {Topology-aware blending method for implicit heterogeneous porous model design},
journal = {Computer-Aided Design},
volume = {177},
pages = {103782},
year = {2024},
issn = {0010-4485},
doi = {https://doi.org/10.1016/j.cad.2024.103782},
author = {Depeng Gao and Yang Gao and Yuanzhi Zhang and Hongwei Lin},
}

@article{Szemer.2026, 
  year     = {2026}, 
  title    = {A persistent homology-based topological lens on solid oxide cell microstructural characterization}, 
  author   = {Szemer, Maksym and Buchaniec, Szymon and Brus, Grzegorz}, 
  journal  = {Journal of Power Sources}, 
  issn     = {0378-7753}, 
  doi      = {10.1016/j.jpowsour.2025.238875}, 
  pages    = {238875}, 
  volume   = {665}
}

@article{Moussaoui.2018, 
  year     = {2018}, 
  title    = {Stochastic geometrical modeling of solid oxide cells electrodes validated on 3D reconstructions}, 
  author   = {Moussaoui, H. and Laurencin, J. and Gavet, Y. and Delette, G. and Hubert, M. and Cloetens, P. and Bihan, T. Le and Debayle, J.}, 
  journal  = {Computational Materials Science}, 
  issn     = {0927-0256}, 
  doi      = {10.1016/j.commatsci.2017.11.015}, 
  pages    = {262--276}, 
  volume   = {143}
}
\end{document}